# Informal specification-based performance evaluation of security protocols

Genge Bela[1], Haller Piroska[1], Iosif Ignat[2] and Ovidiu Răţoi[1]

[1]Department of Electrical Engineering, "Petru Maior" University of Târgu Mureş, Romania,
[2]Department of Computer Science, Technical University of Cluj Napoca, Romania
{bgenge,phaller,oratoi}@engineering.upm.ro, Iosif.Ignat@cs.utcluj.ro

## Abstract

*We propose a performance evaluation method for security protocols. Based on the informal specification, we construct a canonical model which includes, alongside protocol messages, cryptographic operations performed by participants in the process of message construction. Each cryptographic operation is assigned a cost modeled as a function of the size of processed message components. We model not only the size of regular message components but also the size of ciphertext produced by various cryptographic operations. We illustrate the applicability of our method by comparatively analyzing the performance of the original CCITT X.509 protocol and a slightly modified version of the same protocol.*

## 1. Introduction

Security protocols are "communication protocols dedicated to achieving security goals" (C.J.F. Cremers and S. Mauw) [1] such as confidentiality, integrity or availability. Achieving such security goals is made through the use of cryptography.

Designing new protocols is a challenging task if we look at the number of attacks that have been discovered over the years [2] after the protocols have been published. However, in the last few years the use of *protocol composition* [3, 4, 5] has been successfully applied to create new protocols based on existing [6, 7] or predefined protocols [3].

The composition process makes use of the *informal* [6] specification of security protocols which does not include any implementation-related information such as selected cryptographic algorithm, key size or encryption rounds. The result of the composition can have multiple protocols [8] from which the most performant must be selected.

As mentioned earlier, cryptography is an important component of these protocols. This is why existing performance evaluation methods include several aspects related to the performance of the algorithms used to implement the protocols. However, in the composition phase, the cryptographic algorithms used in the implementation process are unknown.

To help the decision process related to the selection of the most performant security protocol, in the early design phase, we propose a novel evaluation method that focuses on cryptographic algorithm operations, available in the informal specification.

The informal specification does not include a formal tool for reasoning on security protocols. In order to achieve our goal, we need such a tool. We have chosen to use the strand space model [9] as a specification model because of its simplicity and extensibility. However, additional information has to be included in the strand-based specification for our goal to be achievable. This is why we first construct an extension of the strand space model, called *k-strand space*, by enriching it with user knowledge and explicit term construction.

The second step is to define the operations that can be extracted from the informal specification and to model them appropriately using a canonical model called *t-strand space*. The canonical model is constructed from the previously defined k-strand model. Because the size of message components of the same type (e.g. random numbers, participant names) is not known, in the t-strand model, message components are constructed from *t-terms* denoting only message types, without any instance-specific values. Based on these terms, we define cryptographic operations used by participants to construct messages.

The proposed performance evaluation method evaluates two protocols against each other. For each protocol, the t-strand model is constructed together with all cryptographic operations needed for participants to construct the messages. Each cryptographic operation is assigned a symbolic cost. The performance of the two protocols is given as the sum of assigned costs and the final values are compared to each other.

The rest of the paper is structured as follows. In Section 2 we present an extension of the original strand space used to model security protocols and we introduce the canonical model where cryptographic operations are modeled as t-strands having specific classifiers. In Section 3 we provide a collection of functions for mapping the cost of each operation and the size of message components. Using the proposed approach, we evaluate the performance of two protocols in Section 4: the CCITT X.509 and a slightly modified version of the same protocol. In Section 5 we relate our work to others found in the literature. We end with a conclusion in Section 6.

## 2. K-strands and t-strands

In this section we briefly present the concept of *knowledge strands* (*k-strands*) and *typed strands* (*t-strands*) used to model protocol participants. For a more detailed presentation, the reader is asked to consult the authors previous work [10].

### 2.1 K-strands and k-strand spaces

A *strand* is a sequence of transmission and reception events used to model protocol participants. A collection of strands is called a *strand space*. The strand space model was introduced by Guttman et all in [9] and extended by the authors with participant knowledge, specialized basic sets and explicit *term* construction in [10]. The resulting model is called a *k-strand space*. The rest of this section formally defines the k-strand and k-strand space concepts.

By analyzing the protocol specifications from the SPORE library [11] we can conclude that protocol participants communicate by exchanging *terms* constructed from elements belonging to the following sets: $R$, denoting the set of participant names; $N$, denoting the set of nonces (i.e. "number once used"); $K$, denoting the set of cryptographic keys and $M$ denoting user-defined components. If required, other sets can be easily added without affecting the other components.

To denote the encryption type used to create cryptographic terms, we define the following *function names*:

$$FuncName ::= sk \quad (secret\ key)$$
$$| pk \quad (public\ key)$$
$$| pvk \quad (private\ key)$$
$$| h \quad (hash)$$

The above-defined basic sets and function names are used in the definition of *terms*, where we also introduce constructors for pairing and encryption:

$$\mathcal{T} ::= . \,|\, R \,|\, N \,|\, K \,|\, M \,|\, (\mathcal{T}, \mathcal{T}) \,|\, \{\mathcal{T}\}_{FuncName(\mathcal{T})},$$

where the '.' symbol is used to denote an empty term. We use the symbol $\mathcal{T}^*$ to denote the set of all subsets of terms.

To denote the transmission and reception of terms, we use *signed terms*. The occurrence of a term with a positive sign denotes transmission, while the occurrence of a term with a negative sign denotes reception. The set of transmission and reception sequences is denoted by $(\pm \mathcal{T})^*$.

**Definition 1.** *A* k-strand (i.e. knowledge strand) *is a tuple* $\langle \mathcal{K}, r, s \rangle$*, where* $\mathcal{K} \in \mathcal{T}^*$ *denotes the knowledge corresponding to the modeled participant,* $r \in R$ *denotes the participant name and* $s \in (\pm \mathcal{T})^*$ *denotes the sequence of transmissions and receptions. A set of k-strands is called a k-strand space. The set of all k-strand spaces is denoted by* $\Sigma_k$*. Let* $\varsigma_k$ *be a k-strand space and* $s_k \in \varsigma_k$ *a k-strand, then:*

1. *We define the following mapping functions: $kknow(s_k)$ to map the knowledge component; $kpart(s_k)$ to map the name component; $kseq(s_k)$ to map the term sequence component;*

2. *A* node *is any transmission or reception of a term, written as* $n = \langle kseq(s_k), i \rangle$*, where i is an integer satisfying the condition* $1 \le i \le klength(s)$*. We define the* $kstrand(n)$ *function to map the k-strand corresponding to a given node;*

3. *Let* $n_1 = \langle kseq(s_k), i \rangle$ *and* $n_2 = \langle kseq(s_k), i+1 \rangle$ *be two consecutive nodes from the same k-strand. Then, there exists an edge* $n_1 \Rightarrow n_2$ *in the same k-strand;*

4. *Let* $n_1, n_2$ *be two nodes. If* $n_1$ *is a positive node and* $n_1$ *is a negative node and* $kstrand(n_1) \ne kstrand(n_2)$*, then there exists an edge* $n_1 \rightarrow n_2$*.*

Figure 1 shows an example graphical specification of Lowe's BAN Concrete Secure RPC [2] protocol in the described k-strand space model. The protocol is modeled as the k-strand space $\varsigma_k = \{s_{kA}, s_{kB}\}$, where

$$s_{kA} = \langle \{A,B,N_a,K_{AB}\}, A, \langle +(A,N_a), -\{N_a,K,B\}_{sk(K_{AB})},$$
$$+\{N_a\}_{sk(K)}, -N_b \rangle \rangle \text{ and}$$
$$s_{kB} = \langle \{B,A,N_b,K_{AB},K\}, B, \langle -(A,N_a),$$
$$+\{N_a,K,B\}_{sk(K_{AB})}, -\{N_a\}_{sk(K)}, +N_b \rangle \rangle.$$

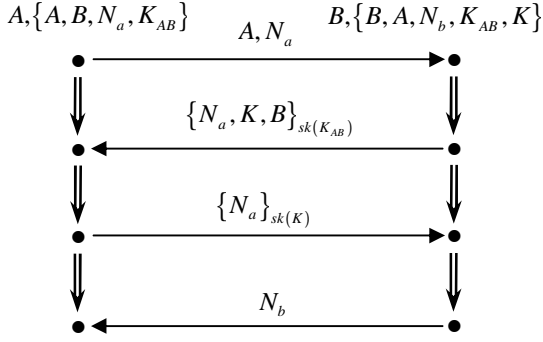

**Figure 1. Lowe's BAN Concrete Andrew Secure RPC representation in the k-strand space model**

## 2.2 T-strands and t-strand spaces

In this section we present the t-strand space model used in the following sections, which is a slightly modified version of the one proposed by the authors in [9]. The model from [9] is enriched with t-strand *classifiers* to denote cryptographic operations corresponding to protocol participants. Thus, protocol participants are modeled using a collection of t-strands.

As opposed to k-strands, in the t-strand model the terms exchanged between t-strands are based on *types* constructed from *basic typed terms* and are called *typed terms* or more simply *t-terms*. Formally, *basic typed terms* and *typed terms* are defined as:

$$
\begin{aligned}
BasicTT &::= r & (participants) \\
&\mid n & (random\ numbers) \\
&\mid k & (keys) \\
&\mid m & (user\text{-}defined)
\end{aligned}
$$

$$\mathcal{T}_t ::= . \mid BasicTT \mid (\mathcal{T}_t, \mathcal{T}_t) \mid \{\mathcal{T}_t\}_{FuncName}$$

By using the above-defined syntactical components we model the instance-independence of security protocols which is a key aspect in the performance evaluation methods that follow in the next sections.

Before defining the concept of typed strands we need to define another element: *classifiers*. As suggested by their names, classifiers are used to classify or categorize typed strands. The categories are created based on the type of operation modeled by a given typed strand. Formally, classifiers are defined as:

$$
\begin{aligned}
C ::=\ & C_{\mathcal{R}} & (Process) \\
\mid\ & C_{\mathcal{E}} & (Symmetric\ encrypt) \\
\mid\ & C_{\mathcal{D}} & (Symmetric\ decrypt) \\
\mid\ & C_{\mathcal{H}} & (Hash) \\
\mid\ & C_{\mathcal{PK}} & (Public\ key\ encrypt) \\
\mid\ & C_{\mathcal{PVK}} & (Private\ key\ encrypt) \\
\mid\ & C_{\mathcal{K}} & (Generate\ key) \\
\mid\ & C_{\mathcal{N}} & (Generate\ nonce) \\
\mid\ & C_{C} & (Concatenate) \\
\mid\ & C_{I} & (Split)
\end{aligned}
$$

Classifiers such as encrypt, decrypt, hash, key and nonce generation, concatenation and split are given to typed strands denoting cryptographic operations specific to protocol participants. On the other hand, the process classifier is given to typed strands used to denote the operations that link together all other classes of typed strands. In other words, process t-strands model the handling of t-terms and cryptographic operations.

A *typed strand* intuitively defines a sequence of t-term transmissions and receptions. Similarly to the k-strand model, we introduce $(\pm \mathcal{T}_t)^*$ to denote the set of t-term transmission and reception sequences. *T-strands* are defined as follows:

**Definition 2.** *A* t-strand *(i.e. typed strand)* *is a tuple* $\langle c,r,s \rangle$, *with* $c \in C$, $r \in R$ *and* $s \in (\pm \mathcal{T}_t)^*$. *A set of t-strands is called a t-strand space. The set of all t-strand spaces is denoted by* $\Sigma_t$. *Let* $\varsigma_t$ *be a t-strand space and* $s_t \in \varsigma_t$ *a t-strand, then:*

1. *We define the following mapping functions:* $tclass(s_t)$ *to map the classifier component;* $tpart(s_t)$ *to map the name component;* $tseq(s_t)$ *to map the t-term sequence component;*
2. *A t-node is any transmission or reception of a t-term, written as* $n_t = \langle tseq(s_t), i \rangle$, *where i is an integer satisfying the condition*

$1 \leq i \leq tlength(s)$. *We define the $tstrand(n)$ function to map the t-strand corresponding to a given t-node;*

3. *Let $n_1 = \langle tseq(s_t), i \rangle$ and $n_2 = \langle tseq(s_t), i+1 \rangle$ be two consecutive t-nodes from the same t-strand. Then, there exists an edge $n_1 \Rightarrow n_2$ in the same t-strand;*

4. *Let $n_1, n_2$ be two t-nodes. If $n_1$ is a positive t-node and $n_1$ is a negative t-node and $tstrand(n_1) \neq tstrand(n_2)$, then there exists an edge $n_1 \rightarrow n_2$.*

In Figure 2 we have the t-strand graphical representation of the protocol from Figure 1. Here, we have intentionally ommitted the representation of cryptographic operations (e.g. encryption, decryption), which will be discussed in the next section. The corresponding t-strand model is $\varsigma_t = \{s_{tA}, s_{tB}\}$, where

$s_{tA} = \langle C_{\mathcal{R}}, A, \langle +(r,n), -\{n,k,r\}_{sk}, +\{n\}_{sk}, -n \rangle \rangle$ and
$s_{tB} = \langle C_{\mathcal{R}}, B, \langle -(r,n), +\{n,k,r\}_{sk}, -\{n\}_{sk}, +n \rangle \rangle$.

### 2.3 Modeling cryptographic operations

Usually, protocol specifications do not include cryptographic operations such as term concatenation, encryption or signature generation, which are considered to be implementation-specific. However, when dealing with the performance evaluation of these protocols we can not omit such operations because they directly influence the evaluation process.

By using the defined classifiers and typed strands, we can model cryptographic operations as follows.

**Definition 3.** *Let $c \in C$ be a classifier. Then the typed strands corresponding to this classifier generate the following sequences of transmissions and receptions for any $t_t, t'_t \in \mathcal{T}_t$:*

*Encryption.* $\langle -t_t, +\{t_t\}_{sk} \rangle$, *if $c = C_{\mathcal{E}}$;*

*Decryption.* $\langle -\{t_t\}_{sk}, +t_t \rangle$, *if $c = C_{\mathcal{D}}$;*

*Hash.* $\langle -t_t, +\{t_t\}_h \rangle$, *if $c = C_{\mathcal{H}}$;*

*Public key enc.* $\langle -t_t, +\{t_t\}_{pk} \rangle$, *if $c = C_{\mathcal{PK}}$;*

*Private key enc.* $\langle -t_t, +\{t_t\}_{pvk} \rangle$, *if $c = C_{\mathcal{PVK}}$;*

*Key-Gen.* $\langle +k \rangle$, *if $c = C_{\mathcal{K}}$;*

*Nonce-Gen.* $\langle +n \rangle$, *if $c = C_{\mathcal{N}}$;*

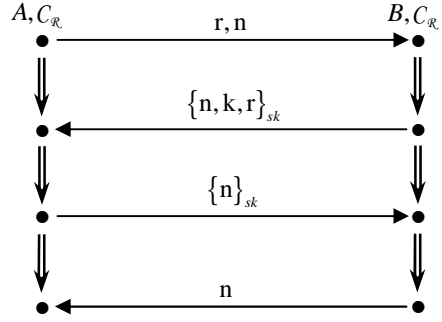

**Figure 2. Lowe's BAN Concrete Andrew Secure RPC representation in the t-strand space model**

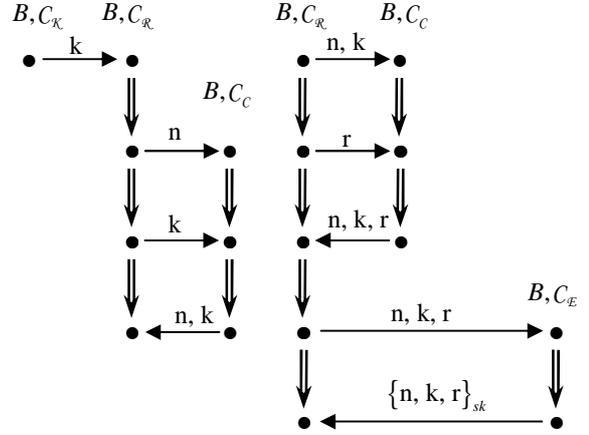

**Figure 3. Encrypted type term generation model**

*Concatenation.* $\langle -t_t, -t'_t, +(t_t, t'_t) \rangle$, *if $c = C_C$;*

*Split.* $\langle -(t_t, t'_t), +t_t, +t'_t \rangle$, *if $c = C_I$.*

Executing a sequence of cryptographic operations produces the terms needed to construct other terms. For example, if we consider the specification given in Figure 2, for the first typed term to be transmitted, a random number has to be generated, followed by the concatenation of two typed terms. The second typed term, $\{n,k,r\}_{sk}$, requires a key generation, the concatenation of three t-terms and the application of a secret key-based encryption function. The graphical representation of the complete construction of the $\{n,k,r\}_{sk}$ t-term is shown in Figure 3.

### 3. Performance evaluation functions

Using the k-strand model, the operations that must be executed by protocol participants are extracted and the t-strand model is constructed. The extraction process uses the knowledge corresponding to each k-strand to identify operations. Thus, terms that are not in the participant's knowledge must be generated (i.e. keys, random numbers) or extracted from encrypted terms which are also located in the knowledge set.

In this paper we consider that for each positive node (i.e. sent terms), terms are extracted from knowledge. If they do not exist, they are generated or extracted from other encrypted terms. In the case of negative nodes (i.e. received terms), we consider that terms are placed in the participant's knowledge. Decryption of terms available in the knowledge set is executed only when needed (i.e. on sending terms). However, participants may need to verify received terms, thus execute decryption operations on every negative node. This requires additional information to be added to the t-strand model, which we consider to be part of a future work.

In the t-strand model, the t-nodes responsible for creating new t-terms have a positive sign. Thus, we assign a cost to each positive t-node found in a t-strand space.

We define the following functions to denote the cost of each operation:

$$f_{sk}, f_{pk}, f_h, f_{kg}, f_{ng}, f_s, f_p : \mathbb{R}^+ \to \mathbb{R}^+$$
$$f_c : \mathbb{R}^+ \times \mathbb{R}^+ \to \mathbb{R}^+,$$

where $f_{sk}$ denotes symmetric key-based operations, $f_{pk}$ denotes public and private key-based operations, $f_h$ denotes hash operations, $f_{kg}$ denotes key generation operations, $f_{ng}$ denotes nonce generation operations, $f_s$ denotes split operations, $f_p$ denotes processing operations and $f_c$ denotes concatenation operations. Each function takes as argument the size of the t-term to be processed and returns the cost of each operation. The $f_{kg}$ and $f_{ng}$ functions receive the size of the t-term to be created. The $f_c$ function receives two arguments: the size of the two t-terms to be concatenated.

Term size plays an important role in the performance of security protocols. Because of this, we define several functions to map the size of t-terms to symbolic values.

The size of each basic typed term is mapped to a symbolic value using the $|\_| : BasicTT \to \mathbb{R}^+$ operator. By applying cryptographic operations on t-terms, the size of the result depends on the type of the algorithm

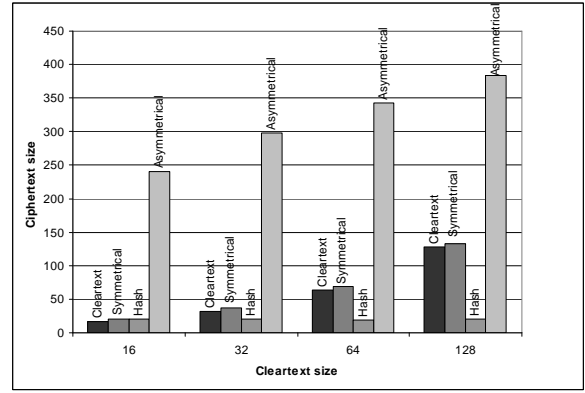

**Figure 4. Cleartext and ciphertext size for symmetrical, hash and asymmetrical algorithms**

that is used [12] (i.e. symmetrical, asymmetrical or hash). In order to model the resulting size, we conducted an exhaustive measurement of the ciphertext size resulted by applying cryptographic algorithms on cleartext of various size. The implementations were chosen from two well-known cryptographic libraries: Cryptlib [13] and OpenSSL [14].

From the results shown in Figure 4 we can clearly state that the size of ciphertext resulting from hash operations is the same for any cleartext while the size of ciphertext resulting from symmetrical cryptographic operations follows the size of cleartext. In contrast, the size of ciphertext resulting from asymmetrical operations is much greater than the original cleartext and strongly depends on the length of the key used in the process [15].

We introduce the $\lambda_S, \lambda_A, \lambda_H : \mathcal{T}_t \to \mathbb{R}^+$ functions to map the size of ciphertext resulted by applying symmetrical, asymmetrical and hash operations respectively, on t-terms. We also introduce the $\Delta : \mathcal{T}_t \to \mathbb{R}^+$ function to map the size of concatenated t-terms. These functions are defined inductively as:

$$\Delta(t) = |t|, \text{ if } t \in BasicTT,$$
$$\Delta(t) = \Delta(t_1) + \Delta(t_2), \text{ if } t = (t_1, t_2),$$
$$\Delta(t) = \lambda_S(t_1), \text{ if } t = \{t_1\}_{sk},$$
$$\Delta(t) = \lambda_A(t_1), \text{ if } t = \{t_1\}_f \wedge f \in \{pk, pvk\},$$
$$\Delta(t) = \lambda_H(t_1), \text{ if } t = \{t_1\}_h$$
$$\lambda_S(t) = \Delta(t),$$
$$\lambda_A(t) = S_{Asym}(\Delta(t)),$$
$$\lambda_H(t) = S_{Hash},$$

where $S_{Hash} \in \mathbb{R}^+$ is a constant denoting the size of hash ciphertext and $S_{Asym} : \mathbb{R}^+ \to \mathbb{R}^+$ is a function that maps the size of asymmetric ciphertext based on the size of cleartext.

## 4. Example performance evaluation

As an example on the usage of our proposed evaluation method, we use the CCITT X.509 one message protocol proposed by M. Abadi and R. Needham [16] as a recommendation for the CCITT.X.509 standard.

The k-strand graphical representation of the protocol is given in Figure 6. We kept the original term notations for clarity, such that $T_a, N_a \in \mathsf{N}$, $X_a, Y_a \in \mathsf{M}$ and $A, B \in \mathsf{R}$. $PK_A, PK_B, SK_A, SK_B \in \mathsf{K}$ are asymmetric keys corresponding to the public and private keys of the two participants.

The performance of the protocol from Figure 6 is evaluated against another protocol with a slightly modified structure. We modify the term signed with the private key of participant A, which becomes

$$\{T_a, h(N_a, B, X_a, \{Y_a\}_{pk(PK_B)})\}_{sk(SK_A)},$$

while the rest of the message remains the same.

Next, by using the t-strand model of the two protocols we prove that the slight modification results in a much less performant protocol. Because of space considerations, we only construct and analyze the t-term containing the differences between the two protocols. The remaining t-terms do not influence the results because the overhead is identical for both protocols.

The graphical representation of the constructed t-strand corresponding to the operations executed by participant *A* in the first protocol is given in Figure 7. The operations executed by participant *B* are the inverse of these.

The cost of the operations executed by participant *A* in the original CCITT X.509 protocol is the sum of the costs of individual operations. For example, the cost of the public key operation in the t-strand with classifier $C_{\mathcal{PK}}$, denoted by a positive t-node, is $f_{pk}(\Delta(m))$ and the cost of the private key operation in the t-strand with classifier $C_{\mathcal{PVK}}$ also denoted by a positive t-node is $f_{pk}(\Delta(\{n,n,r,m,\{m\}_{pk}\}_h)) = f_{pk}(S_{Hash})$.

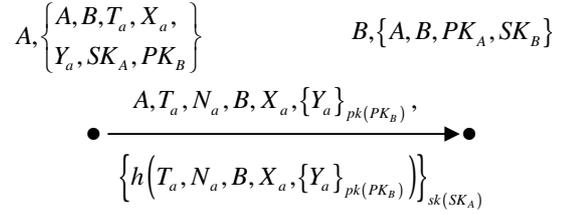

$$A, \begin{cases} A, B, T_a, X_a, \\ Y_a, SK_A, PK_B \end{cases} \qquad B, \{A, B, PK_A, SK_B\}$$

$$A, T_a, N_a, B, X_a, \{Y_a\}_{pk(PK_B)},$$
$$\bullet \xrightarrow{\{h(T_a, N_a, B, X_a, \{Y_a\}_{pk(PK_B)})\}_{sk(SK_A)}} \bullet$$

**Figure 6. K-strand representation of the CCITT X.509 protocol**

The cost of all operations performed by participant *A* is

$Cost_A =$
$\quad f_{pk}(\Delta(m)) +$
$\quad f_{ng}(\Delta(n)) +$
$\quad f_c(\Delta(n), \Delta(n)) +$
$\quad f_c(\Delta((n,n)), \Delta(r)) +$
$\quad f_c(\Delta((n,n,r)), \Delta(m)) +$
$\quad f_c(\Delta((n,n,r,m)), \Delta(\{m\}_{pk})) +$
$\quad f_h(\Delta(n,n,r,m,\{m\}_{pk})) +$
$\quad f_{pk}(\Delta(\{n,n,r,m,\{m\}_{pk}\}_h)) +$
$\quad f_p(\Delta(m)) + 2f_p(\Delta(n)) +$
$\quad f_p(\Delta((n,n))) + f_p(\Delta((n,n,r))) +$
$\quad f_p(\Delta((n,n,r,m))) + f_p(\Delta((n,n,r,m,\{m\}_{pk}))) +$
$\quad f_p(\Delta(\{n,n,r,m,\{m\}_{pk}\}_h)).$

Because of the reduced complexity of the concatenation and processing operations, we consider that the cost is the same for any concatenated or processed t-terms. We use the $\Lambda_C \in \mathbb{R}^+$ symbol to denote the cost of concatenation operations and the $\Lambda_P \in \mathbb{R}^+$ symbol to denote the cost of processing operations. Thus, the cost becomes:

$Cost_A =$
$\quad f_{pk}(|m|) + f_{ng}(|n|) + 4\Lambda_C +$
$\quad f_h(2|n| + |r| + |m| + S_{Asym}(|m|)) +$
$\quad f_{pk}(S_{Hash}) + 8\Lambda_P.$

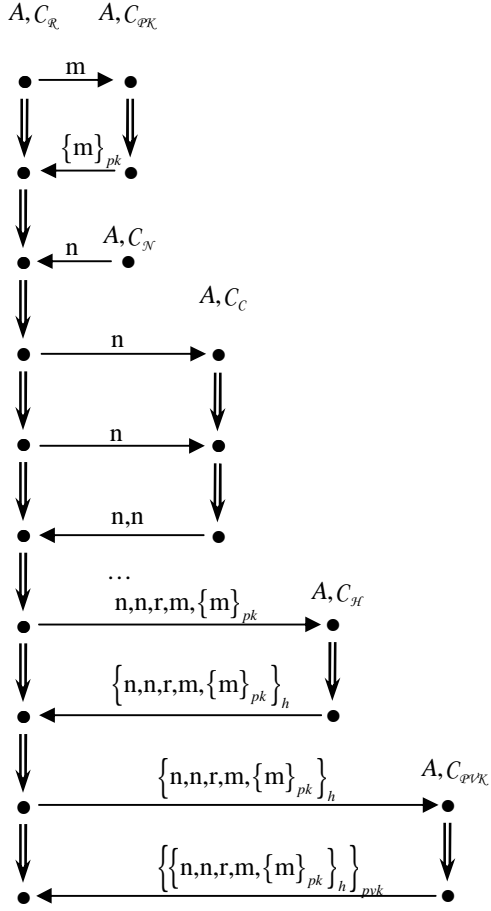

**Figure 7. Encrypted type term generation model**

The cost of the modified CCITT.X protocol is

$$Cost'_A =$$
$$f_{pk}(|m|) + f_{ng}(|n|) + 4\Lambda_C +$$
$$f_h(|n| + |r| + |m| + S_{Asym}(|m|)) +$$
$$f_{pk}(|n| + S_{Hash}) + 8\Lambda_P.$$

In order to decide if the cost of the first protocol is smaller than the cost of the second protocol, we need to establish if $Cost_A < Cost'_A$, meaning that

$$f_{pk}(S_{Hash}) + f_h(2|n| + |r| + |m| + S_{Asym}(|m|)) <$$
$$f_h(|n| + |r| + |m| + S_{Asym}(|m|)) + f_{pk}(|n| + S_{Hash}).$$

By examining the performance of security protocol algorithms from the literature [21], we conclude that

$$f_{alg}(\Delta(t)) = f_{alg}(\Delta(t_1)) + f_{alg}(\Delta(t_2)) - Ovh,$$

where $t = (t_1, t_2)$, $alg \in FuncName$ and $Ovh$ is the overhead of preparing each term for cryptographic operations. The value of the overhead is specific to each cryptographic operation.

Thus, the equation becomes

$$f_h(|n|) - Ovh_H < f_{pk}(|r|) - Ovh_A.$$

If we ignore the overhead, which, compared to the actual cryptographic operation is insignificant, we can clearly state that the cost of the first protocol is smaller than the cost of the second protocol. This is why signature functions are usually applied on the hash of terms and not on the terms themselves.

In the given example, the cost of the two protocols was evaluated against the operations performed by participant *A*. Because the given protocol is a two party protocol, the result would have been the same if we have also evaluated the cost of the operations performed by participant *B*.

## 5. Related work

In the literature we find several papers dealing with the performance evaluation of protocol implementations [17, 18]. In contrast, only a few papers are dedicated to constructing a model for the evaluation of security protocol performance [6, 12].

For completeness, we first mention a few papers that adopted the performance evaluation of various cryptographic algorithm and security protocol implementations. In [17] and [18], the performance of cryptographic algorithms is measured as a function of the total amount of energy consumed by the device on which the algorithm is running. For evaluating the performance of the WTLS [20] (Wireless Transport Layer Security) protocol, the authors from [21] measure the time needed to perform connections on a PDA. Finally, we mention the currently world wide adopted security protocol, TLS [19] (Transport Layer Security). The performance of TLS has been intensively studied [22, 23]. The results show that the cryptographic overhead introduced by TLS seriously affects the performance of regular servers. Because of this, several solutions have been proposed to improve server performance, from which we mention the distribution of cryptographic operations among other servers [23] and the use of hardware accelerators [24].

One of the papers dedicated to modeling the behavior of protocol components [12] constructs a parametric mathematical model based on an exhaustive evaluation process of algorithm implementations. The constructed model does not address, however, the issue of protocol cryptographic operations executed by participants.

A similar approach to ours is proposed in [6] where cryptographic operations are detailed and each operation is assigned a symbolic cost. Our approach differs by the fact that it introduces the concept of size based on term types, as opposed to instance values used in [6]. In addition, we also model the size of message components resulting from cryptographic operations, which is not covered in [6].

## 6. Conclusion

We have developed a procedure for evaluating the performance of security protocols. Our proposal is based on a canonical model which eliminates terms specific to protocol instantiations, leaving only types. The canonical model also includes cryptographic operations that must be executed by protocol participants in order to construct new terms. The total cost associated to cryptographic operations denotes the performance of the analyzed security protocol.

The novelty of our approach lies in the use of participant knowledge to construct cryptographic operations, which does not need any user intervention and provides a minimal effort from participants to create protocol messages. Another novelty introduced by our approach is the association of typed terms to symbolic sizes and the modeling of ciphertext size resulting from cryptographic operations.

As future work, we intend to introduce additional cryptographic operations denoting the verification of received terms. We also intend to use the proposed performance evaluation method in the composition process, which has been used as a method for designing new security protocols from existing protocols. Thus, designers could chose from an early stage the most performant protocol.

## 7. References

[1] C. Cremers, S. Mauw, "Checking secrecy by means of partial order reduction", In S. Leue and T. Systa, editors, Germany, september 7-12, 2003, revised selected papers LNCS, Vol. 3466, 2005, Springer.
[2] Gavin Lowe, "Some new attacks upon security protocols", In Proceedings of the 9th Computer Security Foundations Workshop, IEEE Computer Society Press, 1996, pp. 162-169.
[3] Hyun-Jin Choi, "Security protocol design by composition", Cambridge University, UK, Technical report Nr. 657, UCAM-CL-TR-657, ISSN 1476-2986, 2006.
[4] Ran Canetti, "Universally composable security: A new paradigm for cryptographic protocols", 42nd FOCS, 2001, Revised version (2005), available at eprint.iacr.org/2000/067.
[5] Cas J. F. Cremers, "Compositionality of Security Protocols: A Research Agenda", Electr. Notes Theor. Comput. Sci., 142, pp. 99-110, 2006.
[6] A. Datta, A. Derek, J. C. Mitchell, A. Roy, "Protocol Composition Logic (PCL)", Electronic Notes in Theoretical Computer Science Volume 172, 1 April, 2007, pp. 311-358.
[7] S. Andova, Cas J.F. Cremers, K. Gjosteen, S. Mauw, S. Mjolsnes, and S. Radomirovic, "A framework for compositional verification of security protocols", Elsevier, to appear, 2007.
[8] B. Genge, P. Haller, R. Ovidiu, I. Ignat, "Term-based composition of security protocols", In the Proceeedings of the 16th International Conference on Automation, Quality and Testing, Robotics, AQTR, 2008, to Appear.
[9] F. Javier Thayer Fabrega, Jonathan C. Herzog, Joshua D. Guttman, "Strand spaces: Proving security protocols correct", Journal of Computer Security 7, 1999, pp. 191-230.
[10] Genge Bela, Iosif Ignat, "Verifying the Independence of Security Protocols", IEEE 3rd International Conference on Intelligent Computer Communication and Processing, Cluj-Napoca, Romania, 2007, pp.155-163.
[11] ---, SPORE, Security Protocol Open Repository, http://www.lsv.ens-cachan.fr/spore.
[12] Phongsak Kiratiwintakorn, Energy efficient security framework for wireless Local Area Networks, PhD Thesis, University of Pittsburgh, 2005.
[13] Cryptlib Software Distribution, http://www.cs.auckland.ac.nz/~pgut001/cryptlib/ [last access April, 2008].
[14] OpenSSL Software Distribution, http://www.openssl.org/ [last access April 2008].
[15] Bruce Schneier, Applied Cryptography, John Wiley & Sons, 1996.
[16] Martin Abadi and Roger Needham. Prudent engineering practice for cryptographic protocols, IEEE Transactions on Software Engineering, 22(1), January 1996, pp. 6–15.
[17] M. Viredaz and D. Wallach, "Power evaluation of a handheld computer: A case study," Compaq Western Research Lab, Tech. Rep. 2001/1, 2001.
[18] S. Hirani, "Energy efficiency of encryption schemes in wireless devices", Master's thesis, Telecommunications Program, University of Pittsburgh, Pittsburgh, 2003.
[19] Dierks, T,. Allen, C., "The TLS Protocol, Version 1.0, Request for Comments: 2246", Network Working Group, January 1999.
[20] WAP Forum, "Wireless Transport Layer Security Specification Version 1.1, 11.2.", 1999.
[21] Neil Daswani, "Cryptographic Execution Time for WTLS Handshakes on Palm OS Devices", Certicom Public Key Solutions, September 2000.
[22] Cristian Coarfa, Peter Druschel and Dan S. Wallach, "Performance Analysis of TLS Web Servers", ACM Transactions on Computer Systems, 24 (1), 2006, pp. 39-69.
[23] Adam Stubblefield, Aviel D. Rubin, Dan S. Wallach, Managing the Performance Impact of Web Security, Electronic Commerce Research, No. 5, Springer Science + Business Media, 2005, pp. 99–116.
[24] D. Dean, T. Berson, M. Franklin, D. Smetters, and M. Spreitzer, "Cryptology as a network service", In Proceedings of the 7th Network and Distributed System Security Symposium, San Diego, California, Feb. 2001.